\documentclass[graybox]{svmult}

\usepackage{mathptmx}       
\usepackage{helvet}         
\usepackage{courier}        
\usepackage{type1cm}        
\usepackage{makeidx}         
\usepackage{graphicx}        
\usepackage{multicol}        
\usepackage[bottom]{footmisc}

\makeindex             

\usepackage{amsmath}
\usepackage{amsfonts}
\usepackage{amssymb}
\def\Rl{\mathbb{R}}
\def\Pr{\mathbb{P}}
\def\Ex{\mathbb{E}\,}
\def\Ind{\mathbb{I}}
\def\X{\mathcal{X}}

\def\var{{\rm Var\hskip 0.5pt}}
\def\cov{{\rm Cov}}

\def\med{{\rm med\hskip 0.5pt}}

\def\eps{\varepsilon}
\def\rh{\varrho}

\def\th{\theta}

\def\d{{\rm d}}

\def\dx{\d x}
\def\dy{\d y}

\def\limn{{\lim_{n\to\infty}}}

\def\hth{\hat\theta}
\def\hthI{\hat\theta_{t,n}^{\rm fix}}
\def\hthII{\hat\theta_{r}^{\rm reg}}
\def\hthIII{\hat\theta_{n}^\text{\rm reg-seq}}
\def\hthnu{\tilde\theta_{r}^{\rm unb}} 
\def\hthn{\hat\theta_{n}} 
\def\hthp{\hat\theta_{r}^{\rm perf}} 
\def\fc{\bar{f}}
\def\asvar{\sigma_{\rm as}^{2}(f)}
\def\unvar{\sigma_{\rm unb}^{2}(f)}
\def\tvar{\sigma_{\tau}^{2}}
\def\sitau{\sigma_{\tau}}
\def\sigas{\sigma_{\rm as}}
\def\n0{C_{0}}

\def\lancuch{(X_n)_{n\geq 0}}
\def\stany{\mathcal{X}}
\def\mch{Markov chain}

\newtheorem{thm}{Theorem}[section]
\newtheorem{ass}[thm]{Assumption}

\newtheorem{cor}[thm]{Corollary}
\newtheorem{ex}[thm]{Example}

\begin{document}

\title*{Nonasymptotic bounds on the mean square error
for MCMC estimates via renewal techniques}
\titlerunning{Nonasymptotic bounds for MCMC}
\author{Krzysztof {\L}atuszy\'{n}ski, B{\l}a\.zej Miasojedow and Wojciech Niemiro}
\institute{Krzysztof {\L}atuszy\'{n}ski \at Department of Statistics\\
University of Warwick\\ CV4 7AL, Coventry, UK, \email{latuch@gmail.com}
\and B{\l}a\.zej Miasojedow \at Institute of Applied Mathematics and Mechanics\\ University of Warsaw\\ 
Banacha 2, 02-097 Warszawa, Poland, \email{bmia@mimuw.edu.pl}
\and Wojciech Niemiro \at Faculty of Mathematics and Computer Science\\ Nicolaus Copernicus University\\ 
Chopina 12/18, 87-100 Toru\'{n}, Poland, \email{wniemiro@gmail.com}}
%
%
\maketitle


\abstract{The Nummellin's split chain construction allows to decompose a Markov chain Monte Carlo (MCMC) trajectory into i.i.d. ``excursions''. Regenerative MCMC algorithms based on this technique use a random number of samples. They
have been proposed as a promising alternative to usual fixed length simulation
\cite{Myk95, Ros95JASA, HoJoPrRos}. In this note we derive nonasymptotic bounds on the mean square error (MSE) of regenerative MCMC estimates via techniques of renewal theory and sequential statistics. These results 
are applied to construct confidence intervals. We then focus on two cases of particular interest: chains satisfying the Doeblin condition and a geometric drift condition. Available explicit nonasymptotic results are compared 
for different schemes of MCMC simulation.}

\section{Introduction} \label{sec_intro}

Consider a typical MCMC setting, where $\pi$ is a probability distribution on  $\stany$ and $f:\X\to\Rl$ a 
Borel measurable function. The objective is to compute (estimate) the integral
\begin{equation}\label{eqn_th}
  \th:=\pi f=\int_\X \pi(\dx)f(x).
\end{equation}
Assume that direct simulation from $\pi$ is intractable. Therefore one uses an ergodic Markov chain with transition kernel $P$ and stationary distribution $\pi$ to sample approximately from $\pi$. Numerous computational problems from Bayesian inference, statistical physics or combinatorial enumeration fit into this setting. We refer to \cite{RobRos, RoCa, GilksRichSpiegel} for theory and applications of MCMC.  

Let $\lancuch$ be the Markov chain in question. Typically one discards an initial part of the trajectory 
(called burn-in, say of length $t$) to reduce bias, one simulates the chain for
$n$ further steps and one approximates $\th$ with an ergodic average: 
\begin{equation}\label{eqn_one_walk} 
\hthI = \frac{1}{n}\sum_{i=t}^{t+n-1}f(X_{i}).
\end{equation}
The fixed numbers $t$ and $n$ are the {parameters} of the algorithm.
Asymptotic validity of (\ref{eqn_one_walk}) is ensured by a Strong Law of Large Numbers and a Central Limit Theorem (CLT). Under appropriate regularity conditions \cite{RobRos,BeLaLa}, 
it holds that
\begin{equation}\label{eqn_asnormI} 
\sqrt{n}(\hthI-\th)\to \mathcal{N}(0,\asvar), \qquad(n\to\infty),
\end{equation}
where $\asvar$ is called the asymptotic variance. In contrast with the asymptotic theory, explicit \textit{non\-asymptotic} error bounds for $\hthI$ appear to be very difficult to derive in practically meaningful problems. 

Regenerative simulation offers a way to get around some of the difficulties.
The split chain construction introduced in \cite{AthN78, Num78} (to be described in Section \ref{sec_regeneration}) allows for partitioning the trajectory $\lancuch$ into i.i.d. random tours (excursions) between consecutive regeneration times $T_0, T_1, T_2, \dots$. Random variables 
\begin{equation} \label{eqn_blocks} \Xi_k(f):= \sum_{i= T_{k-1}}^{T_k-1}f(X_i)\end{equation} 
are i.i.d.\ for $k=1,2,\ldots$ ($\Xi_0(f)$ can have a different distribution).
Mykland et al. in \cite{Myk95} suggested a practically relevant recipe for identifying $T_0, T_1, T_2, \dots$ in simulations (formula (\ref{eqn_Mykland}) in Section \ref{sec_regeneration}). This resolves the burn-in problem since one can just ignore the part until the first regeneration $T_0$. One can also stop the simulation at a regeneration time, say $T_{r}$, and simulate $r$ full i.i.d. tours, c.f. Section 4 of \cite{Ros95JASA}. Thus one estimates $\theta$ by 
\begin{equation} \label{eqn_reg_est}
\hthII := \frac{1}{T_{r} -T_0}\sum_{i=T_{0}}^{T_{r}-1}f(X_i)=\frac{\sum_{k=1}^{r}\Xi_k(f)}
{\sum_{k=1}^{r}\tau_k}, 
\end{equation}
where $\tau_{k} = T_k - T_{k-1}=\Xi_k(1)$ are the lengths of excursions.
The number of tours $r$ is fixed and
the total simulation effort $T_{r}$ is random.  
Since $\hthII$ involves i.i.d.\ random variables, classical tools seem to be sufficient to analyse its behaviour.
Asymptotically, (\ref{eqn_reg_est}) is equivalent to (\ref{eqn_one_walk}) because
\begin{equation}\nonumber 
\sqrt{rm}(\hthII-\th)\to \mathcal{N}(0,\asvar), \qquad(r\to\infty),
\end{equation}
where $m := \Ex \tau_1$. Now $rm= \Ex (T_r-T_0)$, the expected length of the trajectory, plays the role of $n$. 
However, our attempt at nonasymptotic analysis in Subsection \ref{sec_est_II} reveals unexpected difficulties:
our bounds involve $m$ in the denominator and  in most practically relevant situations
$m$ is unknown.  

If $m$ is known then instead of (\ref{eqn_reg_est}) one can use an unbiased estimator
\begin{equation}\label{eqn_one_walk_unbiased}
\hthnu:= \frac{1}{rm}\sum_{k=1}^{r}\Xi_k(f),
\end{equation}
Quite unexpectedly, (\ref{eqn_one_walk_unbiased}) is \textit{not equivalent} to (\ref{eqn_reg_est}), even in a weak asymptotic sense. The standard CLT for i.i.d.\ summands yields
\begin{equation}\nonumber 
\sqrt{rm}(\hthnu-\th)\to \mathcal{N}(0,\unvar), \qquad(r\to\infty),
\end{equation}
where $\unvar:={\var \Xi_1(f)}/{m}$ is in general different from $\asvar$.
 
We introduce a new regenerative-sequential simulation scheme, for which  better nonasymptotic results can be derived. Namely, we fix $n$ and define
\begin{equation}\nonumber 
      R(n):=\min\{r: T_{r}> T_0+n\}.
\end{equation}
The estimator is defined as
\begin{equation}\label{eqn_reg_seq_est}
   \hthIII := 
\frac{1}{T_{R(n)}-T_{0}}\sum_{i=T_{0}}^{T_{R(n)}-1}f(X_{i})=\frac{\sum_{k=1}^{R(n)}\Xi_k(f)}
{\sum_{k=1}^{R(n)}\tau_k}. 
\end{equation}
We thus generate a random number of tours as well as a random number of samples. 

Our approach is based on inequalities for the mean square error,
\begin{equation}\nonumber 
 {\rm MSE}:=\Ex(\hth-\th)^2.
\end{equation}
Bounds on the MSE can be used to construct fixed precision confidence intervals. 
The goal is to obtain an estimator $\hth$ which satisfies  
\begin{equation}\label{eq_conf_prec}
 \Pr(|\hth-\th|\leq \varepsilon)\geq 1-\alpha,
\end{equation}
for given $\eps$ and $\alpha$. We combine the MSE bounds with the so called ``median trick'' 
\cite{JeVaViz86, MIS}.  
One runs MCMC repeatedly and computes the median of independent estimates to boost the level of confidence. 
In our paper,  the median trick  is used in conjunction with regenerative simulation. 

 
The organization of the paper is the following. In Section \ref{sec_regeneration} we recall the split chain 
construction. Nonasymptotic bounds for regenerative estimators 
defined by (\ref{eqn_reg_est}), (\ref{eqn_one_walk_unbiased})  and (\ref{eqn_reg_seq_est}) are derived
in Section \ref{sec_est_II_III}.
Derivation of more explicit bounds which involve only computable quantities is deferred to Sections \ref{sec_unif} and \ref{sec_drift}, where we consider classes of chains particularly important in the MCMC context. 
An analogous analysis of the non-regenerative scheme  (\ref{eqn_one_walk})
was considered in \cite{LaNie_rigorous} and (in a different setting and using different
methods) in \cite{Rud}. 

In Section \ref{sec_conf} we discuss the median trick. 
The resulting confidence intervals are compared with \textit{asymptotic} results based on the CLT. 

In Section \ref{sec_unif} we consider Doeblin chains, i.e.\ uniformly ergodic chains that satisfy a one step minorization condition. 
We compare regenerative estimators  (\ref{eqn_reg_est}), (\ref{eqn_one_walk_unbiased})  and (\ref{eqn_reg_seq_est}). 
Moreover, we also consider a perfect sampler available for Doeblin chains,
c.f. \cite{Wilson_read_once, Hob_Rob}. We show that confidence intervals based on the median trick can 
outperform those obtained via exponential inequalities for a single run simulation.

In Section \ref{sec_drift} we proceed to analyze geometrically ergodic Markov chains, 
assuming a drift condition towards a small set. We briefly compare regenerative schemes
(\ref{eqn_reg_est}) and (\ref{eqn_reg_seq_est}) in this setting (the unbiased estimator (\ref{eqn_one_walk_unbiased})  cannot be used, because $m$ is unknown).

\section{Regenerative Simulation}\label{sec_regeneration}

We describe the setting more precisely. Let $\lancuch$ be a Markov chain with transition kernel $P$ on a Polish space $\X$ with stationary distribution $\pi,$ i.e. $\pi P = \pi.$ Assume $P$ is $\pi$-irreducible. 
The regeneration/split construction of Nummelin \cite{Num78} and Athreya 
and Ney \cite{AthN78} rests on the following assumption.

\begin{ass}[Small Set]\label{as: SmallSetCond} 
There exist a Borel set $J\subseteq\X$ of positive $\pi$ measure, a number $\beta>0$ and 
a probability measure $\nu$ such that 
\begin{equation*}
    P(x,\cdot)\geq \beta \Ind(x\in J)\nu(\cdot).
\end{equation*}
\end{ass}

Under Assumption \ref{as: SmallSetCond} we can define a bivariate {\mch}   $(X_n,\Gamma_n)$  
on the space $\X\times\{0,1\}$ in the following way. Variable $\Gamma_{n-1}$ depends only on $X_{n-1}$ via $\Pr(\Gamma_{n-1}=1|X_{n-1}=x)=\beta \Ind(x\in J)$. 
The rule of transition from $(X_{n-1},\Gamma_{n-1})$ to $X_{n}$ is given by 
\begin{equation*}
\begin{split}
  &\Pr(X_{n}\in A|\Gamma_{n-1}=1,X_{n-1}=x)=\nu(A), \\
  &\Pr(X_{n}\in A|\Gamma_{n-1}=0,X_{n-1}=x)=Q(x,A), \\ 
\end{split}
\end{equation*}
where $Q$ is the normalized ``residual'' kernel given by
\begin{equation*}
  Q(x,\cdot):=\frac{P(x,\cdot)- \beta \Ind(x\in J)\nu(\cdot)}{1-\beta \Ind(x\in J)}.
\end{equation*}
Whenever $\Gamma_{n-1}=1$, the chain regenerates at moment $n$.
The regeneration epochs are
\begin{equation*}
\begin{split}
   &T_0:=\min\{n: \Gamma_{n-1}=1\},\\
   &T_k:=\min\{n> T_{k-1}: \Gamma_{n-1}=1\}.
\end{split}
\end{equation*}
The random tours defined by 
\begin{equation} \label{eqn_excursions} \Xi_k : = (X_{T_{k-1}}, \dots, X_{T_{k}-1}, \tau_{k}), \qquad \textrm{where} \quad \tau_{k} = T_k - T_{k-1},
\end{equation}
are independent. Without loss of generality, we assume that $X_{0}\sim\nu(\cdot)$,
unless stated otherwise. Under this assumption, all the tours $\Xi_k$ are  i.i.d.  for $k>0$. We therefore put $T_{0}:=0$ and simplify notation. In the sequel symbols $\Pr$ and $\Ex$ without subscripts refer to the chain started at $\nu$. If the
initial distribution $\xi$ is other than $\nu$, it will be explicitly indicated
by writing $\Pr_{\xi}$ and $\Ex_{\xi}$. Notation $m = \Ex \tau_1$ stands throughout the paper.   

We assume that we are able to \textit{identify} regeneration times $T_{k}$. 
Mykland et al.\ pointed out in \cite{Myk95} that actual sampling from $Q$ can be avoided. We can generate the chain using transition probabability $P$ and then recover the regeneration indicators via
\begin{equation*} \label{eqn_Mykland}
   \Pr(\Gamma_{n-1}=1|X_n,X_{n-1})= 
        \Ind(X_{n-1}\in J)\frac{\beta \nu(\d X_{n})}{P(X_{n-1},\d X_{n})},
\end{equation*}
where $\nu(\dy)/P(x,\dy)$ denote the Radon-Nikodym derivative (in practice, the ratio of densities).
Mykland's trick has been established in a number of practically relevant families (e.g. hierarchical linear models) and specific Markov chains implementations, such as block Gibbs samplers or variable-at-a-time chains, see \cite{JoJo,NeathJones}.

\section{General results for regenerative estimators} \label{sec_est_II_III}

Recall that $f:\X\to\Rl$ is a measurable function and $\th=\pi f$. We consider block sums $\Xi_k(f)$ defined by 
(\ref{eqn_blocks}).
The general Kac theorem states that the mean occupation time 
during one tour is proportional to the stationary measure (Theorem 10.0.1 in \cite{MeynTw}
or Equations (3.3.4), (3.3.6), (3.4.7) and (3.5.1) in \cite{MCMCists}). This yields
\[
m = \frac{1}{\beta \pi(J)}, \qquad \Ex \Xi_1(f) = m \pi f = m\th. \]

From now on we assume that $\Ex \Xi_1(f)^2<\infty$ and $\Ex \tau_1^2<\infty$. For a discussion of these
assumptions in the MCMC context, see \cite{HoJoPrRos}.
Let $\fc := f - \pi f$ and define 
\begin{eqnarray} \label{eqn_as_var}
 \asvar &:= &\frac{\Ex \Xi_1(\fc)^{2}}{m},\\
                 \label{eqn_tau_var} 
  \tvar &:= &\frac{\var \tau_1 }{m}.
\end{eqnarray}
%
\begin{remark}
Under Assumption \ref{as: SmallSetCond}, finiteness of $\Ex \Xi_1(\fc)^{2}$ is a sufficient and necessary condition for the CLT to hold for Markov chain $\lancuch$ and function~$f$. This fact is proved in \cite{BeLaLa} in a more general setting. For our purposes it is important to note 
that $\asvar$ in (\ref{eqn_as_var}) is indeed the \textit{asymptotic variance} which
appears in the CLT.
\end{remark}

\subsection{Results for $\hthII$}\label{sec_est_II}

We are to bound the estimation error which can be expressed as follows:
\begin{equation}\label{eqn_err}  \hthII-\th=\frac{\sum_{k=1}^{r}\big(\Xi_k(f)-\th \tau_k\big)}{\sum_{k=1}^{r}\tau_k}
         =\frac{\sum_{k=1}^{r}d_k}{T_r}.
\end{equation}
where $d_k:=\Xi_{k}(f)-\th \tau_{k}=\Xi_{k}(\fc)$.
Therefore, for any $ 0 < \delta < 1,$
\begin{equation*} 
\Pr (|\hthII - \th| > \varepsilon) \; \leq \; \Pr\left( \bigg|\sum_{k=1}^{r}d_k \bigg|>r m \eps(1-\delta) \right)
         +\Pr\Big( T_r<rm(1-\delta) \Big).
\end{equation*}
Since $d_{k}$ are i.i.d.\  with $\Ex d_{1}=0$ and $\var d_{1}=m\asvar$, we can use Chebyshev inequality to bound the first term above: 
%
\begin{equation}\nonumber 
\Pr\left( \bigg|\sum_{k=1}^{r}d_k \bigg|>r m \eps(1-\delta) \right) \;\leq\; 
\frac{\asvar }{rm\varepsilon^2(1-\delta)^2}.
\end{equation}
The second term can be bounded similarly. We use the fact that  $\tau_{k}$ are i.i.d.\  with $\Ex \tau_{1}=m$ to write 
\begin{equation}\nonumber 
\Pr\Big( T_r<rm(1-\delta) \Big)\;\leq\; \frac{\tvar }{rm^2\delta^2}.
\end{equation}
We conclude the above calculation with in following Theorem.
\begin{thm}\label{thm_conf_bound_general}
Under Assumption \ref{as: SmallSetCond} the following holds for every $0<\delta<1$
\begin{equation}\label{eqn_conf_bound_general}
\Pr (|\hthII - \th| > \varepsilon) \; \leq \;  
\frac{1}{rm}\left[\frac{\asvar }{\varepsilon^2(1-\delta)^2} + \frac{\sigma^2_{\tau} }{m\delta^2}\right]
\end{equation}
and is minimized by \[\delta = \delta^* := \frac{\sitau^{2/3}}{\sigas^{2/3}(f)\eps_{\phantom\tau}^{-2/3}+\sitau^{2/3}}. \]
\end{thm}
Obviously, $\Ex T_r=rm$ is the expected length of trajectory.
The main drawback of Theorem \ref{thm_conf_bound_general} is that the bound on the estimation error depends on $m$,
 which is typically unknown. Replacing $m$ by 1  in \eqref{eqn_conf_bound_general} would be highly inefficient.
This fact motivates our study of another estimator, $\hthIII$,
for which  we can obtain much more satisfactory results. We think
that the derivation of better nonasymptotic bounds  for $\hthII$ (not involving $m$) is an open problem. 

\subsection{Results for $\hthnu$}\label{sec_est_nu}

Recall that $\hthnu$ can be used only when $m$ is known and this situation is rather rare
in MCMC applications. The analysis of $\hthnu$ is straightforward, because it is 
simply a sum of i.i.d.\ random variables. In particular, we obtain the following.
\begin{cor}\label{cor_conf_bound_unbiased} Under Assumption \ref{as: SmallSetCond},
\begin{equation}\nonumber 
\Ex (\hthnu - \th)^2  =  \frac{\unvar}{rm}, \qquad
\Pr (|\hthnu - \th| > \varepsilon) \leq  \frac{\unvar }{rm\,\varepsilon^2}.
\end{equation}
\end{cor} 
Note that  $\unvar={\var \Xi_1(f)}/{m}$ can be
expressed as 
\begin{equation} \label{eqn_un_var}
\unvar =\asvar+\th^2\tvar+2\th\rh(\fc,1),
\end{equation}  
where $\rh(\fc,1):=\cov (\Xi_1(\fc),\Xi_1(1))/m$. 
This follows from the simple observation that $\var \Xi_1(f)=\Ex(\Xi_1(\fc)+\th(\tau_1-m))^2$.
 
\subsection{Results for $\hthIII$}\label{sec_est_III}

The result below bounds the MSE and the expected number of samples used to compute 
the estimator. 
\begin{thm}\label{th: BasicMSE}
If Assumption \ref{as: SmallSetCond} holds then
\begin{equation*}
(i)\quad
   \Ex\,(\hthIII-\th)^2\leq \frac{\asvar}{n^2}\,\Ex\,T_{R(n)}
\end{equation*}
and
\begin{equation*}
(ii)\quad
     \Ex\, T_{R(n)}\leq n + \n0,
\end{equation*}
where
\begin{equation*}
     \n0:=\tvar+m.
\end{equation*}
\end{thm}
\begin{cor}\label{cor_conf_bound_III} Under Assumption \ref{as: SmallSetCond},
\begin{eqnarray} \label{eqn_MSE_bound_III}
 \Ex\,(\hthIII-\th)^2& \leq & \frac{\asvar}{n}\left(1+\frac{\n0}{n}\right), \\
\label{eqn_conf_bound_III}
\Pr\,(|\hthIII-\th|>\eps)& \leq &
     \frac{\asvar}{n\eps^{2}}\left(1+\frac{\n0}{n}\right).
\end{eqnarray}
\end{cor} 
\begin{remark}
Note that the leading term ${\asvar}/{n}$ in (\ref{eqn_MSE_bound_III}) is
``asymptotically correct'' in the sense that the standard fixed length estimator
has ${\rm MSE}\sim {\asvar}/{n}$.  The regenerative-sequential scheme is ``close to the fixed length simulation'',
because $\limn{\Ex T_{R(n)}}/{n}=1$. 
\end{remark}
\begin{proof}[of Theorem \ref{th: BasicMSE}]
Just as in (\ref{eqn_err}) we have
\begin{equation*}
   \hthIII-\th=\frac{\sum_{k=1}^{R(n)}(\Xi_{k}(f)-\th\tau_k)}{\sum_{k=1}^{R(n)}\tau_{k}}
             =\frac{1}{T_{R(n)}}\sum_{k=1}^{R(n)}d_{k},
\end{equation*}
where pairs $(d_{k},\tau_{k})$ are i.i.d.\  with $\Ex d_{1}=0$ and $\var d_{1}=m\asvar$.
Since  $T_{R(n)}> n$, it follows that
\begin{equation*}
\qquad
   \Ex\,(\hthIII-\th)^2\leq 
         \frac{1}{n^{2}}\Ex\left(\sum_{k=1}^{R(n)}d_k\right)^2.
\end{equation*}
Since $R(n)$ is a stopping time with respect to
 ${\cal G}_k=\sigma((d_{1},\tau_{1}),\ldots,(d_{k},\tau_{k}))$, we are in a position to apply the 
two Wald's identities (see Appendix). The second identity yields
\begin{equation*}
\qquad
   \Ex\left(\sum_{k=1}^{R(n)}d_k\right)^2=\var\, d_{1}\,\Ex R(n)=
                   m\asvar\,\Ex R(n).
\end{equation*}
In this expression we can replace $m\Ex R(n)$ by $\Ex T_{R(n)}$ because of the first Wald's identity:
\begin{equation*}
   \Ex\, T_{R(n)}= \Ex \sum_{k=1}^{R(n)}\tau_k=\Ex\tau_{1}\,\Ex R(n)=m\Ex R(n)
\end{equation*}
and (i) follows. 

We now focus attention on bounding the expectation of the ``overshoot'' $\Delta(n):=T_{R(n)}-n$. 
Since we assume that $X_0\sim{\nu}$, the cumulative sums $\tau_1=T_1<T_2<\ldots <T_k<\ldots$ form a (nondelayed) renewal process in discrete time. 
Let us invoke the following elegant theorem of Lorden \cite[Theorem 1]{Lord}:
\begin{equation}\nonumber 
    \Ex \Delta(n) \leq  \Ex \tau_1^2/m.
\end{equation}
This inequality combined with \eqref{eqn_tau_var}
yields immediately $\Ex T_{R(n)}=\Ex (n+\Delta(n))\leq n+\tvar+m$, i.e.\  (ii). 
\end{proof}

\section{The median trick}\label{sec_conf}

This ingeneous method of constructing 
fixed precision MCMC algorithms was introduced in 1986 in  \cite{JeVaViz86}, 
later used in many papers concerned with computational complexity 
and further developed in \cite{MIS}. 
We run $l$ independent copies of the Markov chain. 
Let $\hat\th^{(j)}$ be an estimator computed in $j$th run. 
The final estimate is $\hat\th:=\med(\hat{\th}^{(1)},\ldots,\hat{\th}^{(l)})$.
To ensure that $\hth$ satisfies (\ref{eq_conf_prec}), we require that $\Pr(|\hat\th^{(j)}-\th|>\eps)\leq a$ ($j=1,\ldots,l$) for some modest level
of confidence $1-a<1-\alpha$. This is obtained via Chebyshev's inequality, if a bound on MSE is available.  The well-known Chernoff's inequality gives for odd $l$,
\begin{equation}\label{eq: Chernoff}
\Pr\,(|\hat\th-\th|\geq \eps)\leq \frac{1}{2}\left[4a(1-a)\right]^{l/2}
=\frac{1}{2}\exp\left\{\frac{l}{2}\ln\left[4a(1-a)\right]\right\}.
\end{equation} 
It is pointed out in \cite{MIS} that under some assumptions there is a universal choice of $a$, 
which nearly minimizes the overall number of samples,
$a^{*}\approx  0.11969$. 

Let us now examine how the median trick works in conjunction with regenerative MCMC. We focus on 
$\hthIII$, because  Corollary \ref{cor_conf_bound_III} gives the best available bound on MSE. 
We first choose $n$ such that the right hand side of \eqref{eqn_conf_bound_III} is less than or equal to $a^{*}$. 
Then choose $l$ big enough  to make the right hand side of \eqref{eq: Chernoff} (with $a=a^{*}$) less than or equal to $\alpha$. Compute estimator $\hthIII$ repeatedly, using $l$ independent runs of the chain.
We can see that \eqref{eq_conf_prec} holds if 
\begin{eqnarray}
\label{eq: Suffn}
  n & \geq & \frac{C_1 \asvar}{\eps^{2}}+\n0,\\
\label{eq: Suffl}
  l & \geq & C_2 \ln(2\alpha)^{-1} \;\text{ and $l$ is odd},
\end{eqnarray}
where $C_1:=1/a^{*}\approx  8.3549$ and $C_2:=2/{\ln\left[4a^{*}(1-a^{*})\right]^{-1}}\approx 2.3147$ are absolute constants.
Indeed, \eqref{eq: Suffn} entails $C_1\asvar/(\eps^2n)\leq 1-C_0/n$, so
$C_1\asvar/(\eps^2n)(1+C_0/n)\leq 1-C_0^2/n^2<1$. Consequently $\asvar/(\eps^2n)(1+C_0/n)< a^{*}$ and
we are in a position to apply \eqref{eqn_conf_bound_III}.
 
The overall (expected) number of generated samples is $l\Ex T_{R(n)}\sim nl$ as $\eps\to 0$ and $n\to\infty$,
by Theorem \ref{th: BasicMSE} (ii). Consequently for $\eps\to 0$ the cost of the algorithm 
is approximately 
\begin{equation}\label{eqn_conf_ma} 
   nl\sim {C} \frac{\asvar}{\varepsilon^{2}}\log (2\alpha)^{-1},
\end{equation}
where $C=C_1C_2\approx  19.34$. To see how tight is the obtained lower bound, let us compare 
(\ref{eqn_conf_ma}) with the familiar {asymptotic approximation}, based on the CLT. 
Consider an estimator based on one MCMC run of length $n$, say  $\hthn=\hthI$ with $t=0$. 
From (\ref{eqn_asnormI}) we infer that
\begin{equation*}
  \lim_{\eps\to 0}\; \Pr(|\hthn-\th|> \varepsilon)=\alpha,
\end{equation*}
holds for 
\begin{equation}\label{eqn_conf_asymp} 
  n\sim \frac{\asvar}{\varepsilon^{2}}\left[\Phi^{-1}(1-\alpha/2)\right]^{2},
\end{equation}
where $\Phi^{-1}$ is the quantile function of the standard normal distribution.
Taking into account the fact that $[\Phi^{-1}(1-\alpha/2)]^{2}\sim 2 \log (2\alpha)^{-1}$ for $\alpha\to 0$
we arrive at the following conclusion. The right hand side of \eqref{eqn_conf_ma}
is bigger than \eqref{eqn_conf_asymp} roughly by a constant factor of about 10 
(for small $\eps$ and $\alpha$). The important difference is that \eqref{eqn_conf_ma} 
is sufficient for an \textit{exact} confidence interval while \eqref{eqn_conf_asymp} 
only for an \textit{asymptotic} one.
 
\section{Doeblin Chains} \label{sec_unif}

Assume that the transition kernel $P$ satisfies the following Doeblin condition: there exist
$\beta>0$ and a probability measure $\nu$ such that
\begin{equation}\label{ass_doeblin}
P(x,\cdot) \geq \beta \nu(\cdot) \quad \textrm{for every} \quad x \in \stany.
\end{equation}
This amounts to taking $J:=\stany$ in Assumption \ref{as: SmallSetCond}. Condition
(\ref{ass_doeblin}) implies that the chain is uniformly ergodic. 
We refer to \cite{RobRos} and \cite{MeynTw} for definition of uniform ergodicity and related
 concepts.
As a consequence of the regeneration construction, in our present setting $\tau_1$ is distributed as a geometric random variable with parameter $\beta$ and therefore 
\[m =\Ex \tau_1= \frac{1}{\beta} \qquad \textrm{and} \qquad \sigma_{\tau}^2 = \frac{\var \tau_{1} }{m} =\frac{1-\beta}{\beta}.\]
Bounds on the asymptotic variance $\asvar$ under (\ref{ass_doeblin}) are well known. Let $\sigma^2 = \pi \fc^2$ be the stationary variance. 
Results in Section 5 of \cite{BeLaLa} imply that
\begin{equation}\label{eqn_as_var_unif}
\asvar \leq \sigma^2 \left(1 + \frac{2\sqrt{1-\beta}}{1 - \sqrt{1-\beta}}\right) \leq \frac{4\sigma^2}{\beta}. 
\end{equation}
Since in \cite{BeLaLa} a more general situation is considered, which complicates the formulas, let us give 
a simple derivation of \eqref{eqn_as_var_unif} under \eqref{ass_doeblin}. By  \eqref{eqn_as_var} and
the formula \eqref{eq: SquareBlock} given in the Appendix,
\begin{equation}\nonumber
\asvar \leq  \frac{\Ex \Xi_1(|\fc|)^{2}}{m} \\
       = \Ex_\pi \fc(X_0)^2 + 2\sum_{i=1}^\infty \Ex_\pi |\fc(X_0)\fc(X_i)|\Ind(\tau_1>i). 
\end{equation}
The first term above is equal to $\sigma^2$. To bound the terms of the series,
use Cauchy-Schwarz and the fact that, under \eqref{ass_doeblin}, random variables $X_0$ and $\tau_1$ are 
independent. Therefore $\Ex_\pi |\fc(X_0)\fc(X_i)|\Ind(\tau_1>i)\leq 
\left(\Ex_\pi \fc(X_i)^2 \Ex_\pi \fc(X_0)^2\Pr_\pi(\tau_1>i)\right)^{1/2}=
\sigma^2 (1-\beta)^{i/2}$. Computing the sum of the geometric series yields \eqref{eqn_as_var_unif}.

If the chain is reversible, there is a better bound than  \eqref{eqn_as_var_unif}.  
We can use the well-known formula
for $\asvar$ in terms of the spectral decomposition of $P$ (e.g.  expression ``C'' in \cite{Ros07}). 
Results of \cite{RobRos_electr} show that the spectrum of $P$ is a subset of $[-1+\beta, 1-\beta]$.  
We conclude that for reversible Doeblin chains,
\begin{equation}\label{eqn_as_var_IMH} 
\asvar  \leq \frac{2-\beta}{\beta} \sigma^2\leq \frac{2\sigma^2}{\beta}. 
\end{equation}
An important class of reversible chains are Independence Metropolis-Hastings chains (see e.g. \cite{RobRos}) that are known to be uniformly ergodic if and only if the rejection probability $r(x)$ is uniformly bounded from 1 by say $1- \beta$. This is equivalent to the candidate distribution being bounded below by $ \beta \pi$ (c.f. \cite{MengersenTweedie, AtchadePerron}) and translates into (\ref{ass_doeblin}) with $\nu=\pi$. 
The formula for $\asvar$ in (\ref{eqn_as_var_unif}) and (\ref{eqn_as_var_IMH}) depends on $\beta$ in an optimal way. Moreover (\ref{eqn_as_var_IMH}) is sharp. To see this consider the following example. 
\begin{ex}\label{example_sharp_as_var} Let $\beta \leq  1/2$ and define a Markov chain $(X_n)_{n\geq 0}$ on 
$\mathcal{X} = \{0,1\}$ with stationary distribution $\pi = [1/2, 1/2]$ and transition matrix \begin{equation}\nonumber \qquad P = \left[ \begin{array}{cc} 1-\beta/2 & \beta/2 \\ \beta/2 & 1-\beta/2
\end{array} \right]. \end{equation}
Hence $P = \beta \pi + (1-\beta) I_2$ and $P(x, \cdot) \geq \beta \pi.$ Note that the residual kernel $Q$ is in our example the identity matrix $I_2$.
Thus, before the first regeneration $\tau_1$ the chain does not move. 
Let $f(x) = x.$ Thus $\sigma^2 = 1/4$. To compute $\asvar$ we use
another well-known formula (expression ``B'' in \cite{Ros07}):
\begin{eqnarray*}
\asvar & = & \sigma^2 + 2 \sum_{i=1}^{\infty} \cov_\pi \{f(X_0), f(X_i)\} \\ &=& \sigma^2 + 2\sigma^2 \sum_{i=1}^{\infty} (1-\beta)^i =  \frac{2-\beta}{\beta} \sigma^2.
\end{eqnarray*}   
To compute $\unvar$, note that $\Xi_1(f)=\Ind(X_0=1)\tau_1$.
Since $\tau_1$ is independent of $X_0$ and $X_0\sim \nu=\pi$ we obtain
\begin{eqnarray*}
\unvar  =  \beta\var \Xi_1(f) & = &
                \beta\bigl[\Ex\var (\Xi_1(f)|X_0)+\var \Ex(\Xi_1(f)|X_0)\bigr] \\ 
       & = & \frac{1-\beta}{2\beta}+\frac{1}{4\beta}=\frac{3-2\beta}{\beta}\sigma^2.
\end{eqnarray*} 
Interestingly, in this example $\unvar>\asvar$.
\end{ex}

In the setting of this Section, we will now compare  upper bounds on the total simulation effort needed for different MCMC schemes to get $ \Pr(|\hat\th-\th|>\eps)\leq \alpha$.

\subsection{Regenerative-sequential estimator and the median trick}

Recall that this simulation scheme consists of $l$ MCMC runs, each of approximate length $n$. 
Substituting either (\ref{eqn_as_var_unif}) or (\ref{eqn_as_var_IMH})   in (\ref{eqn_conf_ma}) we obtain that the expected number of samples is 
\begin{equation}\label{eqn_unif_total_cost}
nl \sim 19.34\frac{4\sigma^2}{\beta \eps^2}\log(2\alpha)^{-1} \quad \textrm{and} \quad nl\sim 19.34\frac{(2-\beta)\sigma^2}{\beta \eps^2}\log(2\alpha)^{-1}
\end{equation}
(respectively in the general case and for reversible chains).
Note also that in the setting of this Section we have an exact expression for the constant
$\n0$ in Theorem \ref{th: BasicMSE}. Indeed, $\n0=2/\beta-1$. 

\subsection{Standard one-run average and exponential inequalty}

For uniformly ergodic chains a direct comparison of our approach to exponential inequalities \cite{GlyOr, KontoLastraMeyn} is possible. 
We focus on the result proved in \cite{KontoLastraMeyn} for chains on a countable state space. This inequality is tight in the sense that it reduces 
to the Hoeffding bound when specialised to the i.i.d.\ case.
For $f$ bounded let $\|f\|_{\infty} := \sup_{x\in \mathcal{X}} |f(x)|$. Consider the simple average over $n$ Markov chain samples, say 
$\hthn=\hthI$ with $t=0$.  For an arbitrary initial distribution $\xi$ we have
\begin{equation}\nonumber 
\Pr_{\xi}(|\hthn-\th|>\eps) \leq 2 \exp \left\{ -\frac{n-1}{2}\left(\frac{2\beta}{\|f\|_{\infty}}\eps - \frac{3}{n-1}\right)^2 \right\}.
\end{equation}
After identifying the leading terms we can see that to make the right hand side 
less than $\alpha$ we need 
\begin{equation}\label{eqn_unif_exponential_cost}
n \sim \frac{\|f\|_{\infty}^2}{2\beta^2 \eps^2}\log(\alpha/2)^{-1}\geq \frac{2\sigma^2}{\beta^2 \eps^2}\log(\alpha/2)^{-1}.
\end{equation}
Comparing (\ref{eqn_unif_total_cost}) with (\ref{eqn_unif_exponential_cost}) yields a ratio of roughly 
$40\beta$ or $20\beta$ respectively. This in particular indicates that the dependence on $\beta$ in 
\cite{GlyOr, KontoLastraMeyn} probably can be improved. We note that in examples of practical interest 
$\beta$ usually decays exponentially with the dimension of $\mathcal{X}$ and using the regenerative-sequential-median scheme will often result in a lower total simulation
cost. Moreover, this approach is valid for an unbounded target function $f$, in contrast with 
classical exponential inequalities.

\subsection{Perfect sampler and the median trick}

For Doeblin chains, if regeneration times can be identified, perfect sampling can be performed easily as a version of read-once algorithm \cite{Wilson_read_once}. This is due to the following observation. If condition (\ref{ass_doeblin}) holds and $X_0 \sim \nu$ then 
\begin{equation}\nonumber 
X_{T_k-1}, \quad k = 1,2,\dots  
\end{equation}
are i.i.d. random variables from $\pi$  (see \cite{Breyer_Roberts, MCMCists, Hob_Rob, BeLaLa} for versions of this result). 
Therefore from each random tour between regeneration times one can obtain a single perfect sample 
(by taking the state of the chain prior to regeneration) and use it for i.i.d. estimation. We define
\begin{equation}\nonumber 
\hthp \; := \; \frac{1}{r}\sum_{k=1}^{r} f(X_{T_k-1}).
\end{equation}
Clearly 
\begin{equation}\nonumber 
\Ex (\hthp - \th)^2 \; = \; \frac{\sigma^2}{r}\qquad \textrm{and} \qquad 
\Pr (|\hthp - \th| > \varepsilon) \; \leq \;  \frac{\sigma^2}{r\varepsilon^2}.\;
\end{equation}
Note that to  compute $\hthp$ we need to simulate $n\sim r/\beta$ steps of the Markov chain.
If we combine  the perfect sampler with the median trick we obtain an algorithm with the expected number of samples
\begin{equation}\label{eqn_unif_perfect_cost}
nl\sim 19.34\frac{\sigma^2}{\beta \eps^2}\log(2\alpha)^{-1}.
\end{equation}
Comparing (\ref{eqn_unif_total_cost}) with (\ref{eqn_unif_exponential_cost}) and (\ref{eqn_unif_perfect_cost}) leads to the conclusion that if one targets rigorous nonasymptotic results in the Doeblin chain setting, 
the approach described here outperforms other methods.

\subsection{Remarks on other schemes}

The bound for $\hthII$ in Theorem \ref{thm_conf_bound_general}
is clearly inferior
to that for $\hthIII$ in Corollary \ref{cor_conf_bound_III}.
Therefore we excluded the scheme
based on $\hthII$ from our comparisons.

As for $\hthnu$, this estimator can be used in the Doeblin chains setting, because
$m=1/\beta$ is known. The bounds for $\hthnu$ in Subsection \ref{sec_est_nu} involve $\unvar$.
Although we cannot provide a rigorous proof, we conjecture that in most practical
situations we have $\unvar>\asvar$, because $\rho(\fc,1)$ in (\ref{eqn_un_var}) is
often close to zero. If this is the case, then the bound for $\hthnu$ is inferior 
to that for $\hthIII$. 

\section{A Geometric Drift Condition} \label{sec_drift}

Using drift conditions is a standard approach for establishing geometric ergodicity. 
We refer to \cite{RobRos} or \cite{MeynTw} for the definition and further details. 
The assumption below is the same as in \cite{Bax}. 
Specifically, let $J$ be the small set which appears in Assumption \ref{as: SmallSetCond}. 
\begin{ass}[Drift]\label{as_drift} 
There exist a function $V:\X\to [1,\infty[$, constants $\lambda<1$ and 
$K<\infty$ such that
\begin{equation*}
    PV(x):=\int_{\X}P(x,\d y)V(y)\leq \begin{cases}
               \lambda V(x)& \text{for }x\not\in J,\\
               K& \text{for }x\in J,
              \end{cases}
\end{equation*}
\end{ass}
In many papers conditions similar to Assumption \ref{as_drift} have been established for
realistic MCMC algorithms in statistical models of practical relevance \cite{HoGe98,FoMo,FoMoRoRo,JoHo,JoJo,RoyHo}.
This opens the possibility of computing our bounds in these models.

Under Assumption \ref{as_drift}, it is possible to bound $\asvar$, $\tvar$ and $\n0$ which appear in Theorems \ref{thm_conf_bound_general} and \ref{th: BasicMSE}, by expressions involving only $\lambda$, $\beta$ and $K$.
The following result is a minor variation of Theorem 6.5 in \cite{LaMiaNie}. 
\begin{thm}\label{th: DriftBounds} 
If Assumptions \ref{as: SmallSetCond} and  \ref{as_drift} hold and $f$ is such that 
$\Vert \fc \Vert_{V^{1/2}}:=$ $ \sup_{x}|\fc(x)|/V^{1/2}(x)<\infty$, then  
\begin{equation*}
  \asvar\leq \Vert \fc \Vert_{V^{1/2}}^{2}\left[ \frac{1
+\lambda^{1/2}}{1-\lambda^{1/2}}\pi(V)+\frac{2(K^{1/2}-\lambda^{1/ 2}-\beta(2-\lambda^{1/2}))}{\beta(1-\lambda^{1/ 2})}\pi(V^{1/ 2})\right]
\end{equation*}
\begin{equation*}
    \n0\leq  \frac{\lambda^{1/2} }{1-\lambda^{1/2}} \pi(V^{1/2})
                             +\frac{K^{1/2}-\lambda^{1/2}-\beta}{\beta(1-\lambda^{1/2})}-1.
\end{equation*} 
\end{thm}
To bound $\tvar$ we can use the obvious inequality $\tvar= \n0-m\leq \n0-1$.
Moreover, one can easily control $\pi V$ and $\pi V^{1/2}$ and further replace 
$\Vert \fc \Vert_{V^{1/2}}$ e.g. by $\Vert f \Vert_{V^{1/2}}+ (K^{1/2}-\lambda^{1/2})/(1-\lambda^{1/2})$, we refer to \cite{LaMiaNie} for details.

Let us now discuss possible approaches to confidence estimation in the setting of this section. Perfect sampling is in general unavailable. For unbounded $f$ we cannot apply
exponential inequalities for the standard one-run estimate.
Since $m$ is unknown  we cannot use $\hthnu$.  This leaves $\hthII$ and
$\hthIII$ combined with the median trick. To analyse $\hthII$ we can apply Theorem \ref{thm_conf_bound_general}. Upper bounds for $\asvar$ and $\tvar$ are available.
However, in Theorem \ref{thm_conf_bound_general} we will also need a \textit{lower bound} on $m$. Without further assumptions we can only write 
\begin{equation} \label{eqn_lower_bound_m} m = \frac{1}{\pi(J)\beta} \geq \frac{1}{\beta}.\end{equation}
In the above analysis \eqref{eqn_lower_bound_m} is particularly disappointing. 
It multiplies the bound by an unexpected and substantial factor, as $\pi(J)$ is typically small in applications. 
For $\hthIII$ we have much more satisfactory results.
Theorems \ref{th: BasicMSE} and \ref{th: DriftBounds}  can be used to obtain  bounds which do not involve $m$.
In many realistic examples, the parameters $\beta$, $\lambda$ and $K$ which appear in  Assumptions 2.1 (Small Set) and 6.1 (Drift) 
can be explicitly computed, see e.g.\ \cite{JoHo,JoJo,RoyHo}. 

We note that nonasymptotic confidence intervals for MCMC estimators under drift condition have also been obtained in \cite{LaNie_rigorous}, where 
identification of regeneration times has not been assumed. In absence of regeneration times a different approach has been used and the bounds are typically weaker.
For example one can compare \cite[Corollary 3.2]{LaNie_rigorous} (for estimator $\hthI$) 
combined with the bounds in \cite{Bax} with our Theorems  \ref{th: BasicMSE} and \ref{th: DriftBounds} 
(for estimator $\hthIII$). 
\bigskip\goodbreak

\section*{Appendix}
\addcontentsline{toc}{section}{Appendix}

For convenience, we  recall the two identities of Abraham Wald, which we need in the proof of Theorem \ref{th: BasicMSE}. 
Proofs can be found e.g. in \cite[Theorems 1 and 3 in Section 5.3]{ChowTe}.

Assume that $\eta_1,\ldots,\eta_k,\ldots$, are i.i.d.\ random variables and $R$ is a stopping time such that $\Ex R<\infty$. 
\begin{description}
        \item[{\bf I Wald identity:}] If $\Ex |\eta_1|<\infty$  then 
$$\Ex \sum_{k=1}^{R} \eta_k= \Ex R\, \Ex \eta_1.$$
        \item[{\bf II Wald identity:}] If $\Ex \eta_1=0$ and $\Ex \eta_1^2<\infty$ then 
$$\Ex \left(\sum_{k=1}^{R} \eta_k\right)^2= \Ex R\, \Ex \eta_1^2.$$
\end{description}

In Section \ref{sec_unif} we used the following formula taken from \cite[Equation (4.1.4)]{MCMCists}.  In the notation of our Sections  
\ref{sec_regeneration}  and \ref{sec_est_II_III}, for every $g\geq 0$ we have
\begin{equation}\label{eq: SquareBlock}           
      \frac{\Ex_\nu \Xi_1(g)^{2}}{m}= \Ex_\pi g(X_0)^2+2\sum_{i=1}^\infty \Ex_\pi
                           g(X_0)g(X_i)\Ind(T >i).
\end{equation} 
In \cite{MCMCists} this formula, with $g=\fc$, is used to derive an expression for the asymptotic variance $\asvar=\Ex_\nu \Xi_1(\fc)/m$ 
under the assumption that $f$ is bounded. For $g\geq 0$, the proof is the same.

\end{document}